\documentstyle[aps,prl,twocolumn,graphicx,floats]{revtex}

\begin{document}

\pagenumbering{arabic}
\def\E{{\rm e}}
\newcommand{\be}{\begin{enumerate}}
\newcommand{\ee}{\end{enumerate}}
\newcommand{\beq}{\begin{equation}}
\newcommand{\eeq}{\end{equation}}
\newcommand{\uu}{\underline}
\renewcommand{\thesubsection}{\arabic{subsection}}

\draft
{\wideabs{ 
\title{Partition Function Zeros and Finite Size Scaling\\ of Helix-Coil 
       Transitions in a Polypeptide}
\author{Nelson A. Alves\footnote{alves@quark.ffclrp.usp.br}}
\address{Departamento de F\'{\i}sica e Matem\'atica, FFCLRP 
     Universidade de S\~ao Paulo. Av. Bandeirantes 3900. CEP 014040-901 
     Ribeir\~ao Preto, SP, Brazil}
\author{Ulrich H.E. Hansmann \footnote{hansmann@mtu.edu}}
\address{Department of Physics, Michigan Technological University,
         Houghton, MI 49931-1291, USA}
\date{\today}
\maketitle
\begin{abstract}
We report on multicanonical simulations of the helix-coil transition
of a polypeptide. The nature of this transition was studied by calculating
partition function zeros and the finite-size scaling of various
quantities. New estimates for critical exponents are presented. 
\end{abstract}
}}


A common, ordered structure in proteins is the \hbox{$\alpha$-helix}  and it is 
conjectured that formation of $\alpha$-helices is a key factor in the early 
stages of protein-folding \cite{BSG}. It is long known that $\alpha$-helices 
undergo a sharp transition towards a random coil state when the temperature is
increased. The characteristics of this so-called helix-coil transition have 
been studied extensively \cite{Poland}, 
most recently in Refs.~\cite{Jeff,HO98c}.
They are usually described in the framework of  Zimm-Bragg-type
theories \cite{ZB} in which the homopolymers  are approximated by a
one-dimensional Ising model  with the residues  as ``spins'' taking values 
``helix'' or ``coil'', and solely local interactions. Hence, in such theories
thermodynamic phase transitions are not possible.  However, 
in preliminary work \cite{HO98c} it was shown that our all-atom model of 
poly-alanine  exhibits 
a phase transition between the ordered helical state and the disordered 
random-coil state. It was  conjectured  that this transition 
is due to long range interactions in our
model and  the fact that it is not  one-dimensional:
it is known that the 1D Ising model with long-range interactions
also exhibits  a phase transition at finite $T$ if the interactions
decay like $1/r^{\sigma}$ with $1\le \sigma < 2$ \cite{1dlr}.
Our aim now is to  investigate this transition in the frame work of a critical
theory by means of the finite size scaling (FSS) analysis of partition function 
zeros. Analysis of partition function zeros is a well-known tool in the 
study of phase transitions, but was to our knowledge never  used before to 
study   biopolymers.

For our project, the use of the multicanonical algorithm \cite{MU}
was crucial. The various competing interactions within the polymer lead
to an energy landscape characterized by a multitude of local minima.
Hence, in the low-temperature region, canonical 
simulations will tend to get trapped in one of these
minima and the simulation will not thermalize within the available
CPU time. One standard way to overcome this problem  is the application
of the {\it multicanonical algorithm} \cite{MU} and other 
{\it generalized-ensemble} techniques \cite{Review} to the protein folding 
problem \cite{HO}.  For poly-alanine,  
both the failure of standard Monte Carlo techniques 
and the superior performance of the
multicanonical algorithm are extensively documented  in 
earlier work \cite{OH95b}. 

In the multicanonical algorithm \cite{MU}  
conformations with energy $E$ are assigned a weight
$  w_{mu} (E)\propto 1/n(E)$. Here, $n(E)$ is the density of states.
A  simulation with this weight 
will  lead to a uniform distribution of energy:
\begin{equation}
  P_{mu}(E) \,  \propto \,  n(E)~w_{mu}(E) = {\rm const}~.
\label{eqmu}
\end{equation}
This is because the simulation generates a 1D random walk in the energy,
allowing itself to escape from any  local minimum. 
Since a large range of energies are sampled, one can
use the reweighting techniques \cite{FS} to  calculate thermodynamic
quantities over a wide range of temperatures by
\begin{equation}
<{\cal{A}}>_T ~=~ \frac{\displaystyle{\int dx~{\cal{A}}(x)~w^{-1}(E(x))~
                 e^{-\beta E(x)}}}
              {\displaystyle{\int dx~w^{-1}(E(x))~e^{-\beta E(x)}}}~,
\label{eqrw}
\end{equation}
where $x$ stands for configurations. 

It follows from Eq.~\ref{eqmu} that the multicanonical algorithm allows us
to calculate estimates for the spectral density:
\begin{equation} 
  n(E) = P_{mu} (E) w^{-1}_{mu} (E)~.
\end{equation}
We can therefore construct
the partition function from these estimates by
\begin{equation}
     Z(\beta) = \sum_{E} n(E) u^{E} ,                         \label{eq:r1}
\end{equation}
where $u=e^{-\beta}$ with $\beta$ the inverse temperature, $\beta = 1/k_B T$.
The complex solutions of the partition function  
determine the critical behavior of the model. 
They are the so-called Fisher zeros \cite{fisher,itzykson}, and
correspond to the complex extension of the temperature variable.

Our investigation of the helix-coil transition for poly-alanine is
based on a detailed, all-atom representation of that homopolymer, and
goes beyond the approximations of the Zimm-Bragg model \cite{ZB}.
The interaction between the atoms  was
described by a standard force field, ECEPP/2,\cite{EC}  (as implemented 
in the KONF90 program \cite{Konf}) and is given by:
\begin{eqnarray}
E_{tot} & = & E_{C} + E_{LJ} + E_{HB} + E_{tor},\\
E_{C}  & = & \sum_{(i,j)} \frac{332q_i q_j}{\epsilon r_{ij}},\\
E_{LJ} & = & \sum_{(i,j)} \left( \frac{A_{ij}}{r^{12}_{ij}}
                                - \frac{B_{ij}}{r^6_{ij}} \right),\\
E_{HB}  & = & \sum_{(i,j)} \left( \frac{C_{ij}}{r^{12}_{ij}}
                                - \frac{D_{ij}}{r^{10}_{ij}} \right),\\
E_{tor}& = & \sum_l U_l \left( 1 \pm \cos (n_l \chi_l ) \right).
\end{eqnarray}
Here, $r_{ij}$ (in \AA) is the distance between the atoms $i$ and $j$, and
$\chi_l$ is the $l$-th torsion angle. Note that with the electrostatic
energy term $E_{C}$ our model contains a long range interaction neglected
in the Zimm-Bragg theory \cite{ZB}.
Since one can avoid the
complications of electrostatic and hydrogen-bond interactions of
side chains with the solvent for alanine (a nonpolar amino acid), explicit
solvent molecules were neglected. 
Chains of up to $N=30$ monomers were considered. 
We needed between 40,000 sweeps ($N=10$)  and 500,000 sweeps ($N=30$) for
the weight factor calculations by the iterative  procedure described in 
Refs.~\cite{MU,HO94c}. 
All thermodynamic quantities were  estimated from one
production run of $N_{sw}$ Monte Carlo sweeps starting from a random
initial conformation, i.e. without introducing any bias.
We chose $N_{sw}$=400,000, 500,000, 1,000,000, and 3,000,000 sweeps
for $N=10$, 15, 20, and 30, respectively.

For our analysis of the partition function zeros we first divide the
energy range into intervals of lengths $0.5$ kcal/mol. Equation~\ref{eq:r1}
becomes now a  polynomial in the variable $u$ and can be easily solved
with MATHEMATICA to obtain all complex zeros $u_j^0$ ($j=1,2, ...$).
For the case of $N=10$ we also repeated the calculation of the zeros for
energy bin sizes $1.0$ kcal/mol and $0.25$ kcal/mol. The changes in
the zeros were smaller than the statistical errors. The effect of
the energy bin size on the zeros is also discussed in Ref.~\cite{Kar}.
Figure 1 shows the distribution of the zeros for $N=30$ and  provides 
already strong evidence for a singularity on the real axis: 
in the case of the (analytic) Zimm-Bragg theory 
the zeros would be located  solely on the negative real $u$-axis \cite{Shrock}. 
We summarize in Table I the
leading zeros for each  of the four chain lengths, where we have used the
mapping  $u=e^{-\beta/2}$ due to our binning procedure.

\begin{table}[t]
\begin{tabular}{lllll} 
 \\[-0.4cm] 
 ~$N$ & ~~${\rm Re}\,(u_1^0)$  & ~~${\rm Im}\,(u_1^0)$  & 
  ~~~~${\rm Re}\,(\beta_1^0)$ & ~~${\rm Im}\,(\beta_1^0)$ \\
\\[-0.45cm]
\hline
\\[-0.4cm]
 ~10  & ~$0.5620(60)$  & ~$0.0702(33)$  & ~~~$1.138(21)$  & ~$0.248(11)$ \\
 ~15  & ~$0.6015(23)$  & ~$0.0472(21)$  & ~~~$1.0104(77)$ & ~$0.1566(67)$ \\
 ~20  & ~$0.6105(29)$  & ~$0.03275(88)$ & ~~~$0.9842(94)$ & ~$0.1072(26)$ \\
 ~30  & ~$0.6159(19)$  & ~$0.02200(78)$ & ~~~$0.9681(63)$ & ~$0.0714(25)$ \\
\end{tabular}
\vspace{0.1cm}
\caption{\baselineskip=0.8cm      First partition function zeros for 
          poly-alanine chains of various chain lengths.}
\end{table}

 The FSS relation by Itzykson {\it et al.} \cite{itzykson} for the leading
zero $u_1^0(N)$, 
\beq
  u_1^0(N) = u_c + A N^{-1/ d\nu}[1+O(N^{y/d})]~, ~~~~~~ y<0   \label{eq:r2}
\eeq
shows that the distance from the closest zero $u_1^0$, to the
infinite-chain critical point $u_c = e^{-\beta_c/2}$ on Re($u$)
axis, scales with a relevant linear length $L$, which we translated as 
$N^{1/d}$ in the above equation. Here, $\beta_c$ is the inverse critical
temperature of the infinite long polymer chain and $y$ is the correction
to scaling exponent.
 We remark that, unlike in the Zimm-Bragg model, we have no 
theoretical indication to assume $d$ as a 
particular integer geometrical dimension and report therefore 
 estimates for the quantity $d\nu$.

   For sufficiently large $N$, the exponent $d\nu$ can be obtained
from the linear regression
\beq
-\,{\rm ln}\, |u_1^0(N)-u_c| = \frac{1}{d\nu}\,{\rm ln} (N) + a~. \label{eq:r3}
\eeq
   This relation requires an accurate estimate for $u_c$. Therefore,
 we prefer to calculate  our estimates for $d\nu$ from the corresponding
relation with  $| u_1^0 - u_c|$ replaced by its imaginary part 
${\rm Im}\, u_1^0$.
Including chains of all lengths, $N=10-30$,  
this approach leads to  $d\nu = 0.93(5)$, with a goodness of fit $Q=0.48$. 
Figure~2 displays the corresponding fit. 
Omitting the smallest chain, i.e. restricting the fit to  the range
$N=15-30$, does not change the above result. We obtain now $d\nu=0.93(7)$, 
with $Q=0.22$. This indicates that the $d\nu$ determination is stable 
over the studied chains and therefore, the correction exponent $y$ can be
disregarded in face of the present statistical error.

Considering the real part of the leading zeros given in Table I,
${\rm Re}\, (\beta_1^0(N)) =
 - {\rm ln}\{ [{\rm Re}\,u_1^0(N)]^2\, + [{\rm Im}\,u_1^0(N)]^2 \}$, 
we can 
derive the critical temperature through the following FSS fit 
\cite{fukugita},
\beq {\rm Re}\,(\beta_1^0(N)) = \beta_c + b N^{-1/d\nu} ~.  \label{eq:r4}
\eeq
We obtain $\beta_c = 0.906(12)$ ($Q=0.005$)
for the range $N=10-30$, and $\beta_c =0.929(14)$ ($Q=0.63$) for
$N=15-30$.
The last and more acceptable  estimate, corresponds to 
$T_c(\infty)=541(8)$ K.

       A stronger version for the relation (\ref{eq:r2})  considers that
the next zeros $u_j^0(N)$, should also satisfy a scaling relation 
\cite{itzykson},
\beq
   |u_j^0(N) - u_c| \sim \left( \frac{j}{N} \right)^{1/d\nu}~,
                                                              \label{eq:r6}
\eeq
where $j$ labels the zeros in order of increasing distance from $u_c$.
This relation is expected to be satisfied for large $j$ and allows for an
independent check of the estimate for our exponent $d\nu$. The scaling plot 
in Fig. 3  for the 
roots closest to the critical point $u_c$ demonstrates that the
assumed scaling relation is indeed observed for our data as $N$ increases
and consistent with our estimate of the exponent $d\nu$.

\begin{table}[t]
\begin{center}
\begin{tabular}{llllll} 
\\[-0.4cm] 
 ~$N$ &  ~~~$T_c$ & ~~~$C^{max}$ & ~~$\Gamma_{C}$ &  ~$T_{min}$ & ~$b(T_{min},N)$\\
\\[-0.45cm]
\hline
\\[-0.4cm]
 ~10& ~427(7)& ~~8.9(3)   &  ~150(7) & ~~298 & ~~-0.48(4)\\
 ~15& ~492(5)& ~~12.3(4)  &  ~119(5) & ~~429 & ~~-0.59(10)\\
 ~20& ~508(5)& ~~16.0(8)  &  ~88(5)  & ~~469 & ~~-0.55(8)\\
 ~30& ~518(7)& ~~22.8(1.2)&  ~58(4)  & ~~500 & ~~-0.20(4)\\
\end{tabular}
\end{center}
\caption{\baselineskip=0.8cm
 Numerical results for poly-alanine chains of various lengths:
 critical temperature $T_c$ defined by the maximum of specific heat $C_{max}$,
 width  $\Gamma_{C}$ of peak in specific heat and
 temperature $T_{min}$ where the Binder cumulant $b(T,N)$ 
 has its minimum, $b(T_{min},N)$.}
\end{table}
 Our  results for the critical temperature and critical exponent can be
compared with independent estimates obtained from FSS of the specific heat:
 \begin{equation}
  C_N (T)  = {\beta}^2 \ (<E^2(T)> - {<E(T)>}^2)/N~.
\label{eqsh}
\end{equation}
Defining the critical temperature $T_c(N)$   as the
position  where the specific heat $C_N(T)$ has its maximum,
we can again calculate the critical temperature by means of Eq.~\ref{eq:r4}.
With the values in Table~II we obtain 
$T_c(\infty) = 544(12)$ K, which is   consistent with  
the value obtained from the partition function zeros analysis,
$T_c(\infty) =541(8)$ K.
Choosing   $T_1(N)$ and $T_2(N)$ 
 such that $C(T_1) = 1/2 \,C(T_c) = C(T_2)$,  we  have the 
following scaling relation
for the width $\Gamma_C (N)$ of the specific heat \cite{fukugita},
\begin{equation}
\Gamma_C (N) = T_2(N) - T_1(N) \propto N^{-1/d\nu}.
\label{nu1}
\end{equation}
Using the above equation and the values given in Table~II,
we obtain  $d\nu = 0.98(11)$ $(Q=0.9)$
for chains of length $N=15$ to $N=30$, i.e. omitting the shortest 
chain.
This value is in agreement with 
our estimate $d\nu=0.93(5)$, 
 obtained from the partition function zero analysis. 
Including $N=10$ leads to $d\nu=1.19(10)$, but with a
less acceptable fit ($Q=0.1$). 
The analysis of partition function zeros seems
also to be more stable than one relying on Eq.~\ref{nu1}. 
No significant change in $d\nu$ was observed
when the data from ref.~\cite{HO98c} (which relied on much smaller
number of Monte Carlo sweeps) were used in the partition function zeros 
analysis, while Eq.~\ref{nu1} leads for this reduced
statistics to an estimate for $d\nu=1.9$.

Through the scaling relation for the peak of the specific heat, we can 
evaluate yet another critical exponent, the specific heat exponent 
$\alpha$, by: 
\begin{equation}
C_N^{max} \propto  N^{\alpha/d\nu}~.
\label{alpha}
\end{equation}
In particular, with the values for $C_N^{max}$ as given in Table~II,
we obtain $\alpha = 0.86(10)$. 
The scaling plot for the specific heat is shown in Fig.~4: curves for all
lengths of the poly-alanine chains nicely collapse on each other indicating
the scaling of the specific heat and the reliability of our exponents.
 It worths to note that our estimates for $d\nu$ and $\alpha$,
as obtained from the finite size scaling of the specific heat,
obey within the errorbars the hyperscaling relation
$ d\nu = 2 - \alpha $.

It is well-known that renormalization-group fixed point picture 
leads to a critical exponent $d\nu=1$, $\alpha=1$ and $\gamma =1$
 for a first-order phase transition  
\cite{fukugita,fisher_nu,alves_nu}. 
Our estimate $d\nu = 0.93(5)$ for the correlation exponent  deviates from unity
and rather indicates that the `helix-coil-transition'' is
a strong second order transition. However, the errorbars are such that
a first order phase transition cannot be excluded. Our values 
for the specific heat exponent $\alpha=0.86(10)$ and the
susceptibility exponent $\gamma=1.06(14)$ (data not shown) are  
consistent with a first order phase transition, but also not conclusive.
A common way to evaluate the order of a phase transition is by means of
the Binder energy cumulant \cite{binder1},
\begin{equation}
 b(T,N) = 1 - \frac{<E^4(T,N)>}{3 <E^2(T,N)>^2}~.
\end{equation}
For a second order phase transition one would expect that the 
minimum of this quantity $b(T_{min},\infty)$ approaches $2/3$.
 Here $T_{min}$ defines the temperature where the cumulant 
 reaches its minimum value and
$ b(T_{min},\infty) = \lim_{N \rightarrow \infty} b(T,N) $.
With the present values of Table II we find the infinite 
volume extrapolation
$ b(T_{min},\infty) = 0.23(13)$ $(Q=0.12)$, for the range $N=15-30$,
which is consistent with a first order phase transition.
  However, we cannot exclude the possibility 
of a second order phase transition because the energy cumulant scales with the
maximum of specific heat \cite{berg1}, 
$ b(T,N) \sim N^{\alpha/d\nu -1} $, 
and the true asymptotic limit is reached only for 
rather large chains due to the value of $\alpha/d\nu$. 
In fact,
the straight line fit for the range $N=10-30$ is less consistent with
 our data ($Q\simeq 0.001$). 
Hence, we conclude that our results seem to favor a 
(weak) first order phase transition, but are not precise enough to  exclude
the possibility of a second order phase transition.

To summarize, we have used a  common technique for
investigation of phase transitions,  analysis of the finite size scaling
of partition function zeros, to evaluate the helix-coil 
transition in an all-atom model of poly-alanine. 
 Although our results are due to the complexity of the simulated model 
not precise enough to determine the order of the phase transition, we
 have demonstrated that
the transition can be described by a set of critical exponents.  
 Hence, we have shown for  this example that structural transitions 
in biological molecules  can be described within 
the frame work of a critical theory. 

\noindent
{\bf Acknowledgements}: \\
Financial supports from FAPESP and a Research Excellence
Fund  of the State of Michigan
are gratefully acknowledged. 




\newpage
{\Large Figure Captions:}\\
\begin{enumerate}
\item Partition function zeros in the complex $u$ plane for $N=30$.
      For the Zimm-Bragg model the zeros would be located solely on
      the negative real $u$ axis.
\item Linear regression for -ln\,(Im$\,u^0_1(N)$) in the range $N=10-30$.
\item Scaling behavior of the first $j$ complex zeros closest
      to $u_c = 0.6284$, for chain lengths $N = 10,15,20$ and 30.
\item Scaling plot for the specific heat $C_N(T)$ as a function of 
      temperature $T$, for poly-alanine molecules of chain lengths 
      $N=10, 15, 20,$ and $30$.
\end{enumerate}

\newpage
\cleardoublepage

\begin{figure}[t]
\begin{center}
\begin{minipage}[t]{0.95\textwidth}
\centering
\includegraphics[angle=-90,width=0.72\textwidth]{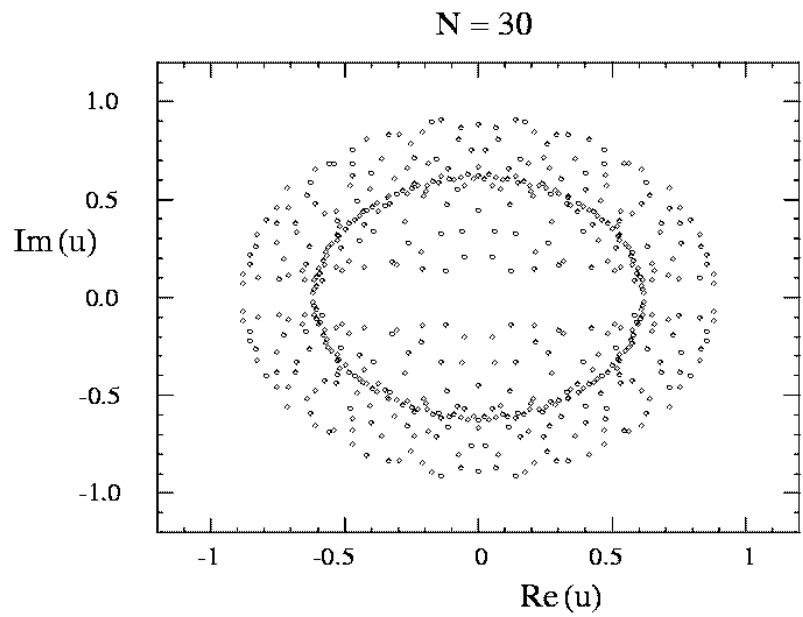}
\renewcommand{\figurename}{FIG.}
\caption{Partition function zeros in the complex $u$ plane for $N=30$.
         For the Zimm-Bragg model the zeros would be located solely on
         the negative real $u$ axis.}
\label{fig1}
\end{minipage}
\end{center}
\end{figure}

\newpage
\cleardoublepage

\begin{figure}[!ht]
\begin{center}
\begin{minipage}[t]{0.95\textwidth}
\centering
\includegraphics[angle=-90,width=0.72\textwidth]{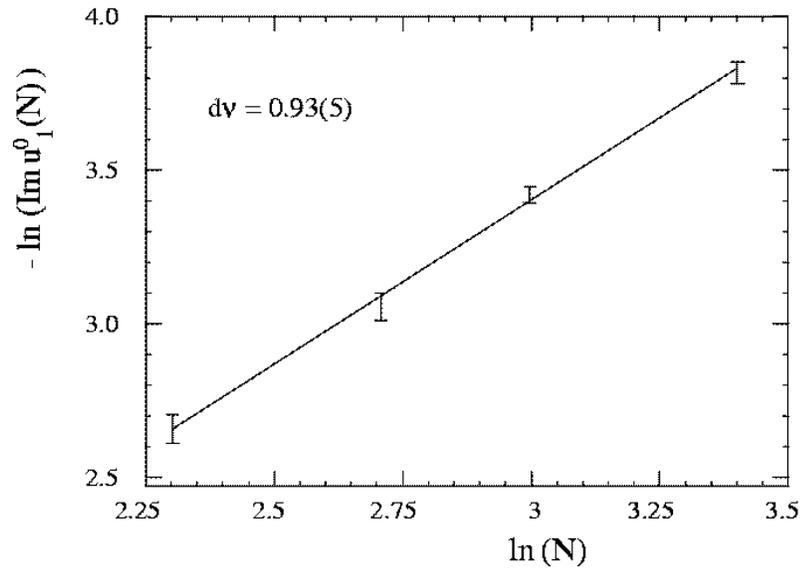}
\renewcommand{\figurename}{FIG.}
\caption{Linear regression for -ln\,(Im$\,u^0_1(N)$) in the range $N=10-30$.}
\label{fig2}
\end{minipage}
\end{center}
\end{figure}

\newpage
\cleardoublepage

\begin{figure}[!ht]
\begin{center}
\begin{minipage}[t]{0.95\textwidth}
\centering
\includegraphics[angle=-90,width=0.72\textwidth]{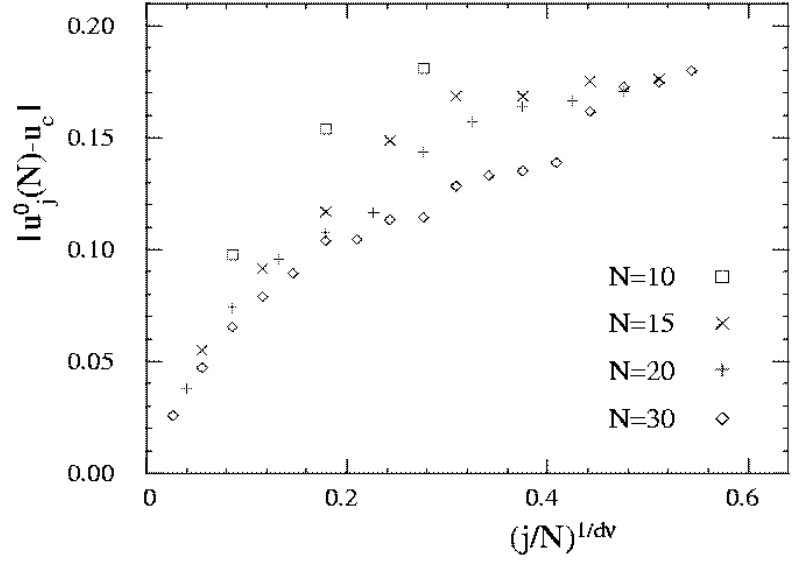}
\renewcommand{\figurename}{FIG.}
\caption{Scaling behavior of the first $j$ complex zeros closest
      to $u_c = 0.6284$, for chain lengths $N = 10,15,20$ and 30.}
\label{fig3}
\end{minipage}
\end{center}
\end{figure}

\newpage
\cleardoublepage

\begin{figure}[t]
\begin{center}
\begin{minipage}[t]{0.95\textwidth}
\centering
\includegraphics[angle=-90,width=0.72\textwidth]{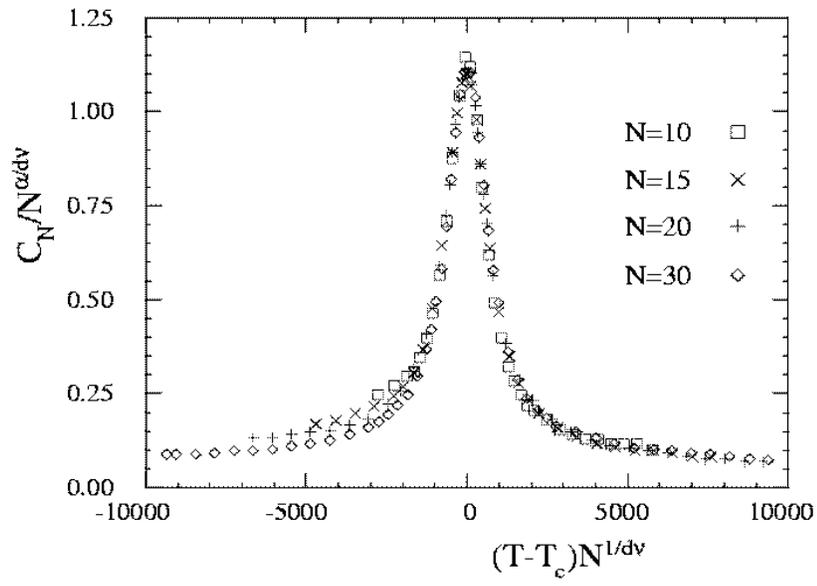}
\renewcommand{\figurename}{FIG.}
\caption{Scaling plot for the specific heat $C_N(T)$ 
         as a function of temperature $T$, for poly-alanine 
         molecules of chain lengths $N = 10, 15, 20$ and 30.}
\label{fig4}
\end{minipage}
\end{center}
\end{figure}

\end{document}